\documentclass[a4paper,11pt]{article}
\pdfoutput=1 

\usepackage{jheppub_modified} 

\usepackage[T1]{fontenc} 

\usepackage{verbatim}
\usepackage{color}
\usepackage{graphicx}
\usepackage{subcaption}

\usepackage{bigcenter}

\usepackage{sectsty}
\allsectionsfont{\boldmath}

\newcommand{\eq}[1]{Eq.~(\ref{#1})}
\newcommand{\bib}[1]{Ref.~\cite{#1}}

\newcommand{\bibs}[1]{\cite{#1}}
\newcommand{\fig}[1]{Fig.~\ref{#1}}
\newcommand{\tab}[1]{Table~\ref{#1}}

\newcommand{\sect}[1]{Section~\ref{#1}}

\newcommand{\bea}{\begin{eqnarray}}
\newcommand{\eea}{\end{eqnarray}}

\newcommand{\crn}{\nonumber \\}

\newcommand{\fr}{\frac}

\def\MG5{{\tt MadGraph5\_aMC@NLO}}

\newcommand{\gev}{{\unskip\,\text{GeV}}}
\newcommand{\tev}{{\unskip\,\text{TeV}}}

\newcommand{\pb}{{\unskip\,\text{pb}}}

\title{Polarization observables in WZ production at the 13 TeV LHC: Inclusive case}

\author[a]{Julien Baglio,}
\author[b]{Le Duc Ninh}

\affiliation[a]{CERN, Theoretical Physics Department, CH-1211 Geneva 23, Switzerland}
\affiliation[b]{Institute For Interdisciplinary Research in Science and Education,\\  
ICISE, 590000 Quy Nhon, Vietnam}

\emailAdd{julien.baglio@cern.ch}
\emailAdd{ldninh@ifirse.icise.vn}

\preprint{CERN-TH-2019-181, IFIRSE-TH-2019-7}      
\abstract{We present a study of the polarization observables of the
  $W$ and $Z$ bosons in the process $p p \to W^\pm Z\to e^\pm \nu_e
  \mu^+\mu^-$ at the 13 TeV Large Hadron Collider. The calculation is
  performed at next-to-leading order, including the full QCD
  corrections as well as the electroweak corrections, the latter being
  calculated in the double-pole approximation. The results are
  presented in the helicity coordinate system adopted by ATLAS and for
  different inclusive cuts on the di-muon invariant mass. We define
  left-right charge asymmetries related to the polarization fractions
  between the $W^+ Z$ and $W^- Z$ channels and we find that these
  asymmetries are large and sensitive to higher-order effects. Similar
  findings are also presented for charge asymmetries related to a
  P-even angular coefficient.}

\begin{document}
\maketitle
\flushbottom

\section{Introduction}
\label{sect:intro}
In the framework of the Standard Model (SM) of particle physics the
$W$ boson only interacts with left-handed fermions while the $Z$ boson
interacts with both left- and right-handed fermions, albeit with
different coupling strengths. This allows for a polarized production
at hadron colliders and in particular at the CERN Large Hadron Collider
(LHC), leading to asymmetries in the angular distributions of the
leptonic decay products of the electroweak gauge bosons. Measuring
these asymmetries is a probe of the underlying polarization of the
gauge bosons and eventually of their spin structure.

The pair production of $W$ and $Z$ bosons has been the subject of
recent experimental studies~\cite{Aaboud:2019gxl} in order to gain
information about the polarization of the gauge bosons. On the theory
side, leading order (LO) studies began a while
ago~\cite{Bilchak:1984gv,Willenbrock:1987xz} before being
revived~\cite{Stirling:2012zt} and studied in a recent
paper~\cite{Baglio:2018rcu} in the process $p p\to W^\pm Z\to e^\pm
\nu_e \mu^+\mu^-$ at next-to-leading order (NLO)
including both QCD and electroweak (EW) corrections, the latter
being calculated in the double-pole approximation (DPA). This
approximation works remarkably well in this process as shown by the
comparison performed in \bib{Baglio:2018rcu} with the exact NLO EW
calculation for the differential
distributions~\cite{Biedermann:2017oae}, which completed the NLO EW
picture after the on-shell predictions presented in
Refs.~\cite{Bierweiler:2013dja,Baglio:2013toa}. Note that for the
production process itself the QCD corrections are known up to
next-to-next-to-leading order in
QCD~\cite{Ohnemus:1991gb,Frixione:1992pj,Grazzini:2016swo,Grazzini:2017ckn}. In
\bib{Baglio:2018rcu} the extensive study of the NLO QCD+EW predictions
for gauge boson polarization observables, namely polarization
fractions and angular coefficients, was done in two different
coordinate systems, the Collins-Soper \cite{Collins:1977iv} and helicity \cite{Bern:2011ie} 
coordinate systems. However, in the recent experimental
analysis by ATLAS with 13 TeV LHC data~\cite{Aaboud:2019gxl}, a
different coordinate system was used, namely a modified helicity
coordinate system in which the $z$ axis is now defined as the
direction of the $W$ (or $Z$) boson as seen in the $WZ$ center-of-mass
system. In addition, the study in \bib{Baglio:2018rcu} introduced
fiducial polarization observables, which have the advantage of being
much simpler to define and calculate (and should also be measurable),
but are not the observables that are measured by the experiments yet.

The goal of this paper is to make one step closer to the experimental
setup by using the modified helicity coordinate system and giving
predictions in an inclusive setup, using as default the experimental total phase space 
defined by ATLAS at the 13 TeV LHC. It is noted that polarization fractions 
at the total-phase-space level are needed in \cite{Aaboud:2019gxl} to simulate 
the helicity templates necessary to extract the polarization fractions
in the fiducial-phase-space region.

In addition to providing results for polarization fractions and angular coefficients, 
we also present two charge asymmetries (denoted $\mathcal{A}^V_{LR}$
and $\mathcal{C}^V_{3}$ with $V=W,Z$) that are large at the NLO QCD+EW
accuracy, which are sensitive to either the QCD or the EW corrections
depending on the asymmetry and on the gauge boson that is under
consideration. These asymmetries help to probe the underlying spin
structure of the gauge bosons and should be measurable in the
experiments. We use the same calculation setup presented in
\bib{Baglio:2018rcu}, and our calculation is exact at NLO QCD using
the program {\tt VBFNLO}~\cite{Arnold:2008rz,Baglio:2014uba} while the
EW corrections are calculated in the DPA presented in
\bib{Baglio:2018rcu}.

Compared to our previous work \cite{Baglio:2018rcu}, the differences in this paper are the
phase-space cuts and the definition of the coordinate system to
determine the lepton angles. Instead of using more realistic fiducial
cuts as in Ref.~\cite{Baglio:2018rcu}, we use here only one simple cut
on the muon-pair invariant mass, e.g. $66\gev < m_{\mu^+\mu^-} <
116\gev$. The benefit of considering this inclusive phase space is
that the ``genuine'' polarization observables can be easily calculated
using the projection method defined in Ref.~\cite{Baglio:2018rcu}. The
word ``genuine'' here means that the observables are not affected by
cuts on the kinematics of the individual leptons such as $p_{T,\ell}$
or $\eta_{\ell}$. This inclusive case may therefore provide more
insights into various effects, which are difficult to understand when
complicated kinematical cuts on the individual leptons are present.

We discuss in \sect{sect:pol} the polarization observables of a
massive gauge boson in the total phase space and our method to
calculate them. The coordinate system that we use to determine the lepton angles is also defined.
In \sect{sect:results} numerical results for the
polarization fractions and angular coefficients are presented for both
$W^+ Z$ and $W^- Z$ channels. From this, results for various charge
asymmetries between the two channels are calculated. We finally
conclude in \sect{sect:conclusion}.

\section{Polarization observables}
\label{sect:pol}
The definition of polarization observables and calculational details
have been all given in Ref.~\cite{Baglio:2018rcu} and we will not
repeat them extensively. For an easy reading of this paper, we provide here a brief summary of
polarization observables and the main calculational details. This will be
needed to understand the numerical results presented in the next
section. Polarization observables associated with a massive gauge
boson are constructed based on the angular distribution of its decay
product, typically a charged lepton (electron or muon).  
In the rest frame of the gauge boson, this distribution
reads~\cite{Collins:1977iv,Chaichian:1981va,Mirkes:1992hu}
\begin{align}
\fr{d\sigma}{\sigma d\!\cos\theta d\phi} &= \fr{3}{16\pi}
\Big[ 
(1+\cos^2\theta) + A_0 \fr{1}{2}(1-3\cos^2\theta)
+ A_1 \sin(2\theta)\cos\phi  \crn
& + A_2 \fr{1}{2} \sin^2\theta \cos(2\phi)
+ A_3 \sin\theta\cos\phi + A_4 \cos\theta \crn
& + A_5 \sin^2\theta \sin(2\phi) 
+ A_6 \sin(2\theta) \sin\phi + A_7 \sin\theta \sin\phi
\Big],\label{eq:definition_Ai_tot}
\end{align}
where $\theta$ and $\phi$ are the lepton polar and azimuthal angles,
respectively, in a particular coordinate system that needs to be
specified. $A_{0-7}$ are dimensionless angular coefficients
independent of $\theta$ and $\phi$. $A_{0-4}$ are called P-even and 
$A_{5-7}$ P-odd according to the parity transformation where $\phi$ 
flips sign while $\theta$ remains unchanged \cite{Hagiwara:1984hi,Frederix:2014cba}. 
We also note here that $A_{5-7}$ are proportional to the imaginary parts 
of the spin-density matrix of the $W$ and $Z$ bosons in the DPA 
at LO \cite{Aguilar-Saavedra:2015yza,Aguilar-Saavedra:2017zkn,Baglio:2018rcu}. 
This is important to understand why the values of these coefficients are very small, as 
will be later shown. 

We can also define the polarization fractions $f^{W/Z}$ by integrating
over $\phi$,
\begin{align}
\fr{d\sigma}{\sigma d\!\cos\theta_{e^\pm}} &= \fr{3}{8}
\Big[ 
(1\mp \cos\theta_{e^\pm})^2 f^{W^{\pm}}_L + (1\pm \cos\theta_{e^\pm})^2 f^{W^{\pm}}_R 
+ 2 \sin^2\theta_{e^\pm} f^{W^{\pm}}_0 \Big],\crn
\fr{d\sigma}{\sigma d\!\cos\theta_{\mu^-}} &= \fr{3}{8} 
\Big[ 
(1 + \cos^2\theta_{\mu^-} + 2c\cos\theta_{\mu^-}) f^{Z}_L + (1 + \cos^2\theta_{\mu^-} - 2c\cos\theta_{\mu^-}) f^{Z}_R
+ 2 \sin^2\theta_{\mu^-} f^{Z}_0 \Big].
\label{eq:def_fLR0}
\end{align}
The upper signs are for $W^+$ and the lower signs are for $W^-$. The
parameter $c$ reads
\bea
c = \fr{g_L^2 - g_R^2}{g_L^2 + g_R^2} =
\fr{1-4s^2_W}{1-4s^2_W+8s^4_W},\quad s^2_W = 1 - \fr{M_W^2}{M_Z^2},
\label{eq:def_c}
\eea
occurring because the $Z$ boson decays into both left- and
right-handed leptons. Relations between the polarization fractions
$f^V_{L,R,0}$ with $V=W,Z$ and the angular coefficients are therefore
obvious,
\begin{align}
f^V_L 
  &= \fr{1}{4}(2- A^V_0 + b_V A^V_4),\;\; f^V_R = \fr{1}{4}(2-
    A^V_0 - b_V A^V_4),\;\;
f^V_0 = \fr{1}{2}A^V_0,\label{eq:relations_fL0R_Ai}
\end{align}
where $b_{W^\pm} = \mp 1$, $b_Z = 1/c$. From this, we get
\bea
f^V_L + f^V_R + f^V_0 = 1, \quad f^V_L - f^V_R = \fr{b_V}{2} A^V_4.
\eea

\begin{figure}[ht!]
  \centering
  \includegraphics[height=5cm]{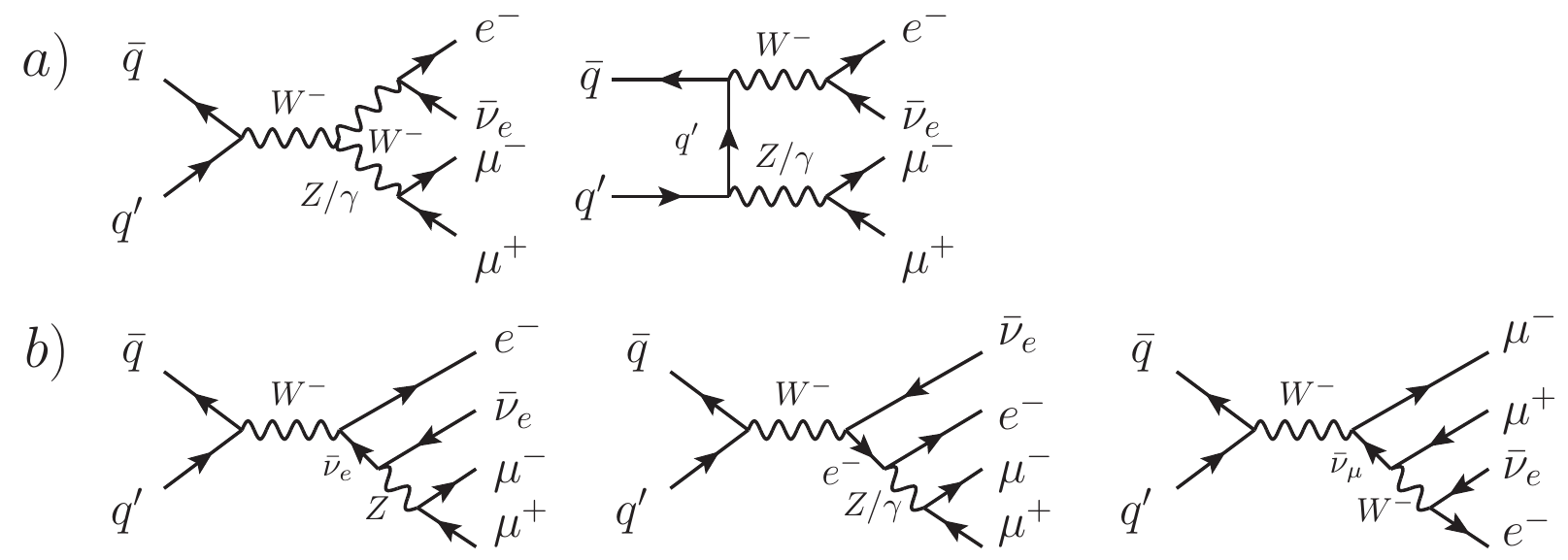}
  \caption{Double and single resonant diagrams at leading order. Group
    a) includes both double and single resonant diagrams, while group
    b) is only single resonant.}
  \label{fig:LO_diags}
\end{figure}

These coefficients are named polarization observables because they are
directly related to the spin-density matrix of the $W$ and $Z$ bosons
in the DPA and at LO as above mentioned. In order to calculate them,
we first have to calculate the distributions $d\sigma/(d\cos\theta
d\phi)$, or simply the distributions $d\sigma/(d\cos\theta)$ if only
the polarization fractions are of interest.
This can be computed order by order in
perturbation theory. We have calculated this up to the NLO QCD + EW
accuracy using the same calculation setup as in \bib{Baglio:2018rcu}. 
The NLO QCD results are exact, using the full amplitudes as
provided by the {\tt VBFNLO}
program. The NLO EW corrections
are however calculated in the DPA as presented in
Ref.~\cite{Baglio:2018rcu}. In the DPA, only the double-resonant
Feynman diagrams are taken into account. Single-resonant diagrams
including $\gamma^* \to \mu^+ \mu^-$ (as shown in
Fig.~\ref{fig:LO_diags}a) or $W\to 2\ell 2\nu$ (as shown in
\fig{fig:LO_diags}b) are neglected. Moreover, even for the
double-resonant diagrams, off-shell effects are not included. In the
next section we will also provide results at LO using the DPA (dubbed
DPA LO) or using the full amplitudes (dubbed simply LO).

Finally, we specify the coordinate system to determine the angle
$\theta$ and $\phi$. Differently from Ref.~\cite{Baglio:2018rcu}, we
use here the modified helicity coordinate system. The only difference
compared to the helicity system is the direction of the $z$ axis:
instead of being the gauge boson flight direction in the laboratory
frame as chosen in Refs.~\cite{Bern:2011ie,Baglio:2018rcu}, it is now
the gauge boson flight direction in the $WZ$ center-of-mass
frame. This modified helicity coordinate system is also used in the
latest ATLAS paper presenting results for the polarization observables
in the $WZ$ channel~\cite{Aaboud:2019gxl}. We think the modified
helicity system is a better choice when studying the spin correlations
of the two gauge bosons. However, for polarizations of a single gauge
boson, the helicity system is more advantageous because of a better
reconstruction of the $Z$ boson direction in the laboratory frame. In
both cases, an algorithm to determine the momentum of
the $W$ boson from its decay products is still needed, which 
has been done in \cite{Aaboud:2019gxl}. We note that the spin correlations of the two gauge bosons are fully included in our calculation. 
However, we do not provide separately numerical results for these effects in this paper because the need for them 
is not urgent as the current experimental-statistic level is still limited to be sensitive to those effects. 
Nevertheless, we choose to use the modified helicity system to be closer to the ATLAS measurement
and to prepare for the future studies of those spin correlations.  

\section{Numerical results}
\label{sect:results}
The input parameters are
\begin{eqnarray}
G_{\mu} = 1.16637\times 10^{-5} \gev^{-2}, \,
M_W=80.385 \gev, \, 
M_Z = 91.1876 \gev, \crn 
\Gamma_W = 2.085\gev, \, \Gamma_Z = 2.4952\gev, \, 
M_t = 173 \gev, \, M_H=125\gev, 
\label{eq:param-setup}
\end{eqnarray}
which are the same as the ones used in \bib{Baglio:2018rcu}. The
masses of the leptons and the light quarks, {\it i.e.} all but the top
mass, are approximated as zero. This is justified because our results
are insensitive to those small masses. The electromagnetic coupling is
calculated as $\alpha_{G_\mu}=\sqrt{2}G_\mu
M_W^2(1-M_W^2/M_Z^2)/\pi$. For the factorization and renormalization scales, we use 
$\mu_F = \mu_R = (M_W + M_Z)/2$. Moreover, the parton distribution functions (PDF) are 
calculated using the Hessian set 
{\tt
  LUXqed17\char`_plus\char`_PDF4LHC15\char`_nnlo\char`_30}~\bibs{Manohar:2016nzj,Manohar:2017eqh,Butterworth:2015oua,Dulat:2015mca,Harland-Lang:2014zoa,Ball:2014uwa,Gao:2013bia,Carrazza:2015aoa,Watt:2012tq,deFlorian:2015ujt} via the library {\tt LHAPDF6}~\bibs{Buckley:2014ana}.

We will give results for the LHC running at a center-of-mass  energy
$\sqrt{s} = 13\,\tev$, for both $e^+\nu_e\,\mu^+\mu^-$ and
$e^-\bar{\nu}_e\,\mu^+\mu^-$ final states, also denoted, respectively, 
as $W^+ Z$ and $W^- Z$ channels for conciseness. We treat the extra
parton occurring in the NLO QCD corrections inclusively and we do not
apply any jet cuts. We also consider the possibility of lepton-photon
recombination, where we redefine the momentum of a given charged
lepton $\ell$ as being $p'_\ell = p_\ell + p_\gamma$ if $\Delta
R(\ell,\gamma) \equiv \sqrt{(\Delta\eta)^2+(\Delta\phi)^2}< 0.1$. We
use $\ell$ for either $e$ or $\mu$. If not otherwise stated, the
default phase-space cut is
\bea
66\gev < m_{\mu^+\mu^-} < 116\gev ,
\label{eq:cut_default}
\eea
which is used in \bib{Aaboud:2016yus,Aaboud:2019gxl} to define the
experimental total phase space. With this cut, we obtain the following 
result for the total cross section
\bea
\sigma^\text{tot.}_{W^\pm Z,\text{NLO QCD+EW}} = 45.8 \pm 0.7\, \text{(PDF)} + 2.2/-1.8\, \text{(scale)}\, \pb ,
\label{eq:xsection_tot}
\eea
where we have used $\text{Br}(W \to e \nu_e) = 10.86\%$ and $\text{Br}(Z\to \mu^+\mu^-) = 3.3658\%$ as provided in \bib{Agashe:2014kda} to unfold the 
cross section as done in \bib{Aaboud:2019gxl}. This result is to be compared with $\sigma^\text{tot.}_{W^\pm Z,\text{ATLAS}} = 51.0 \pm 2.4\, \pb$ as reported in \bib{Aaboud:2019gxl}, 
showing a good agreement at the $1.6\sigma$ level. 
The agreement becomes even much better when the next-to-next-to-leading order QCD corrections, of the order of a $+11\%$ on top of the NLO QCD results at 13 TeV for our scale choice, 
are taken into account~\cite{Grazzini:2016swo}. 
We note that the EW corrections to the total cross section are completely negligible (at the sub-permil level) because of the cancellation between the negative corrections 
to the $\bar{q}q'$ channels and the positive corrections to the $q\gamma$ channels, in agreement with the finding 
in \bib{Baglio:2013toa}.

\subsection{Angular distributions and polarization fractions}
\begin{figure}[ht!]
  \centering
  \begin{tabular}{cc}
  \includegraphics[width=0.48\textwidth]{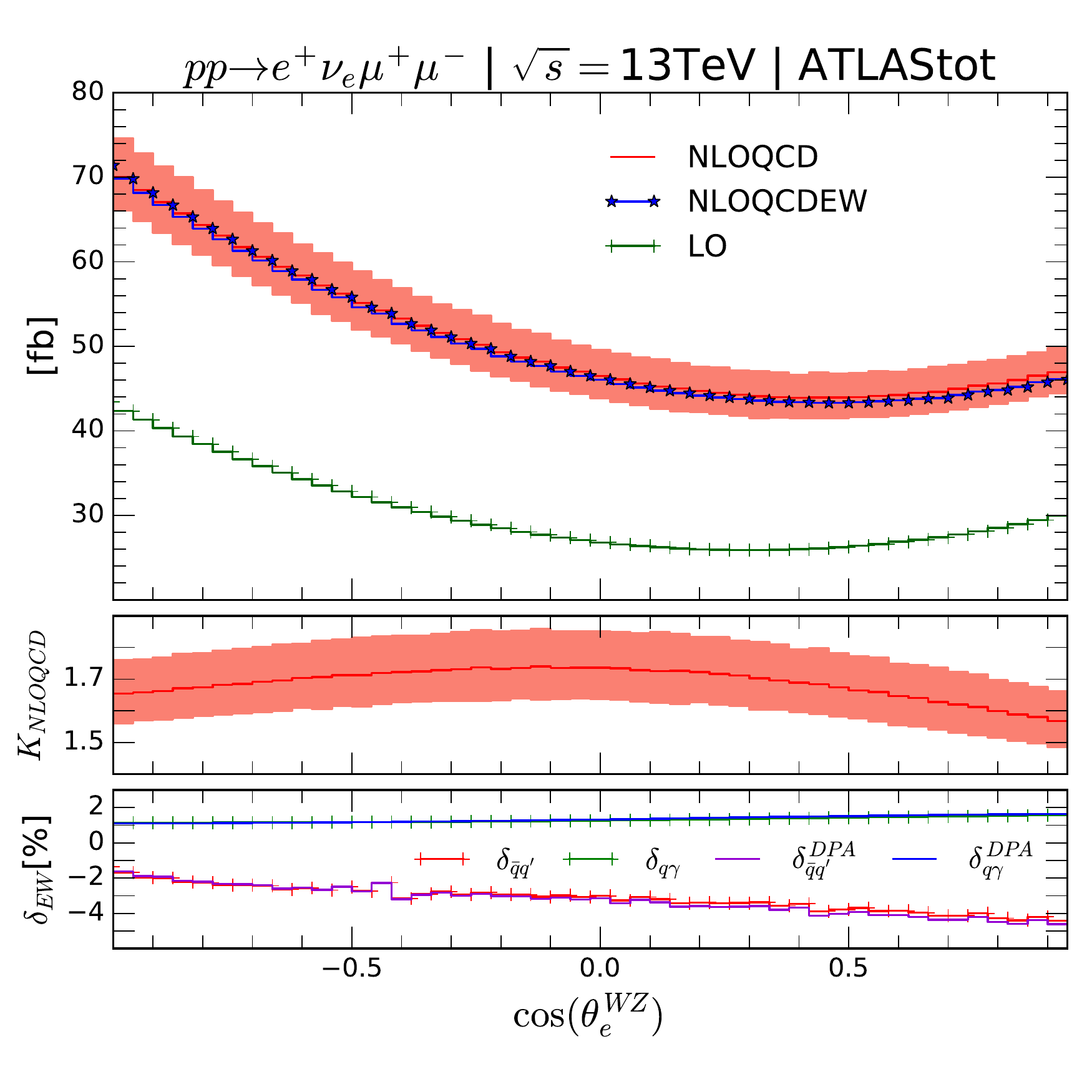}&
  \includegraphics[width=0.48\textwidth]{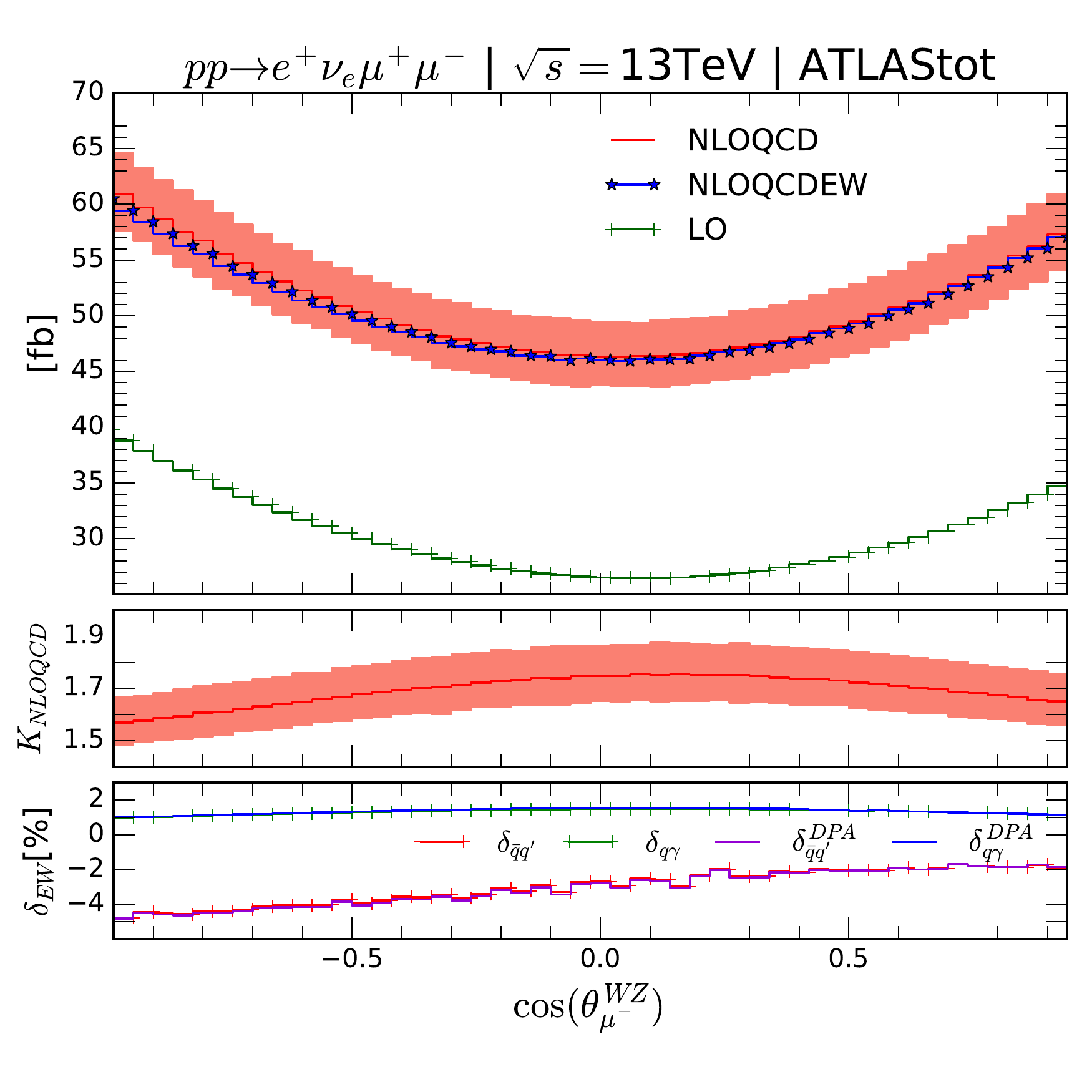}\\
  \includegraphics[width=0.48\textwidth]{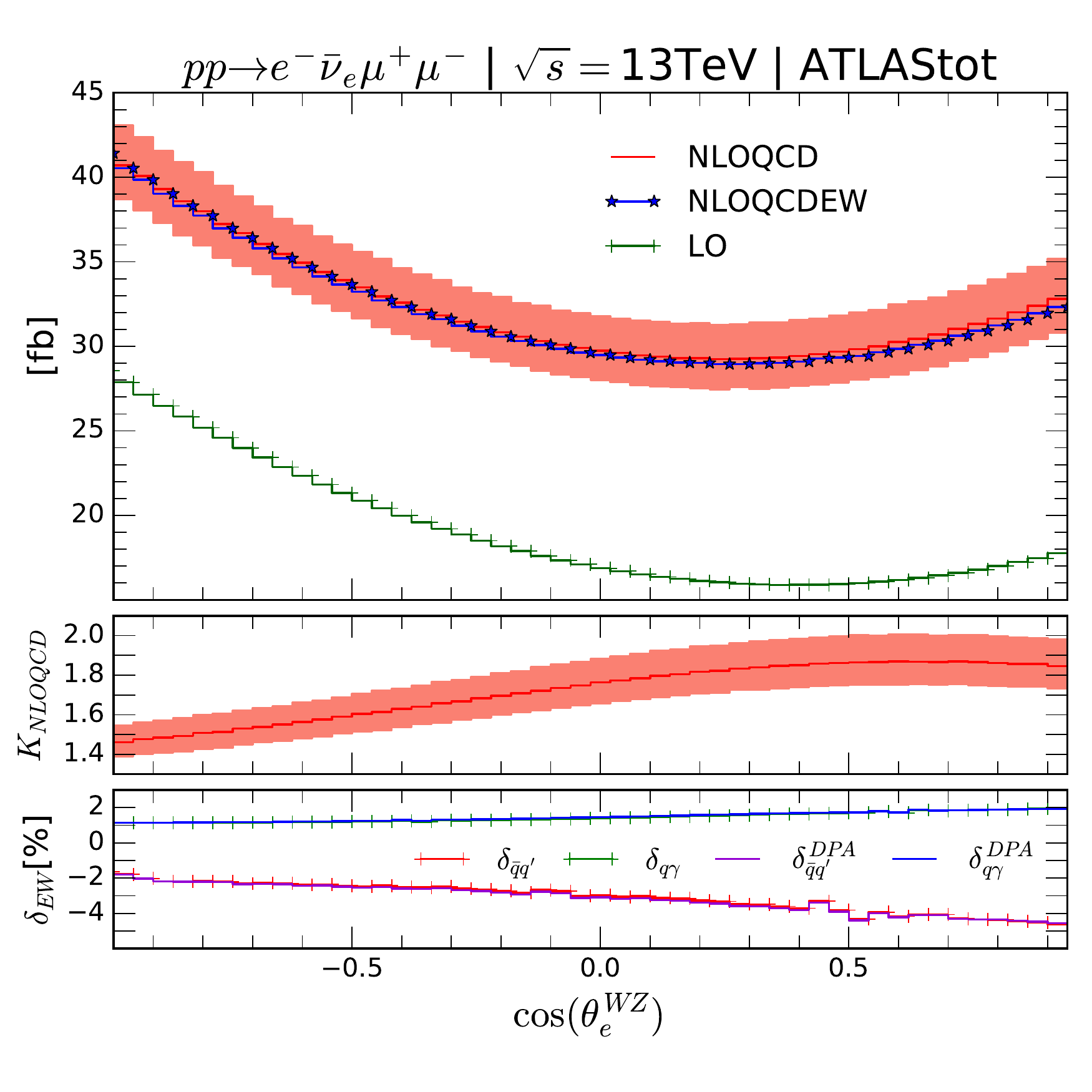}&
  \includegraphics[width=0.48\textwidth]{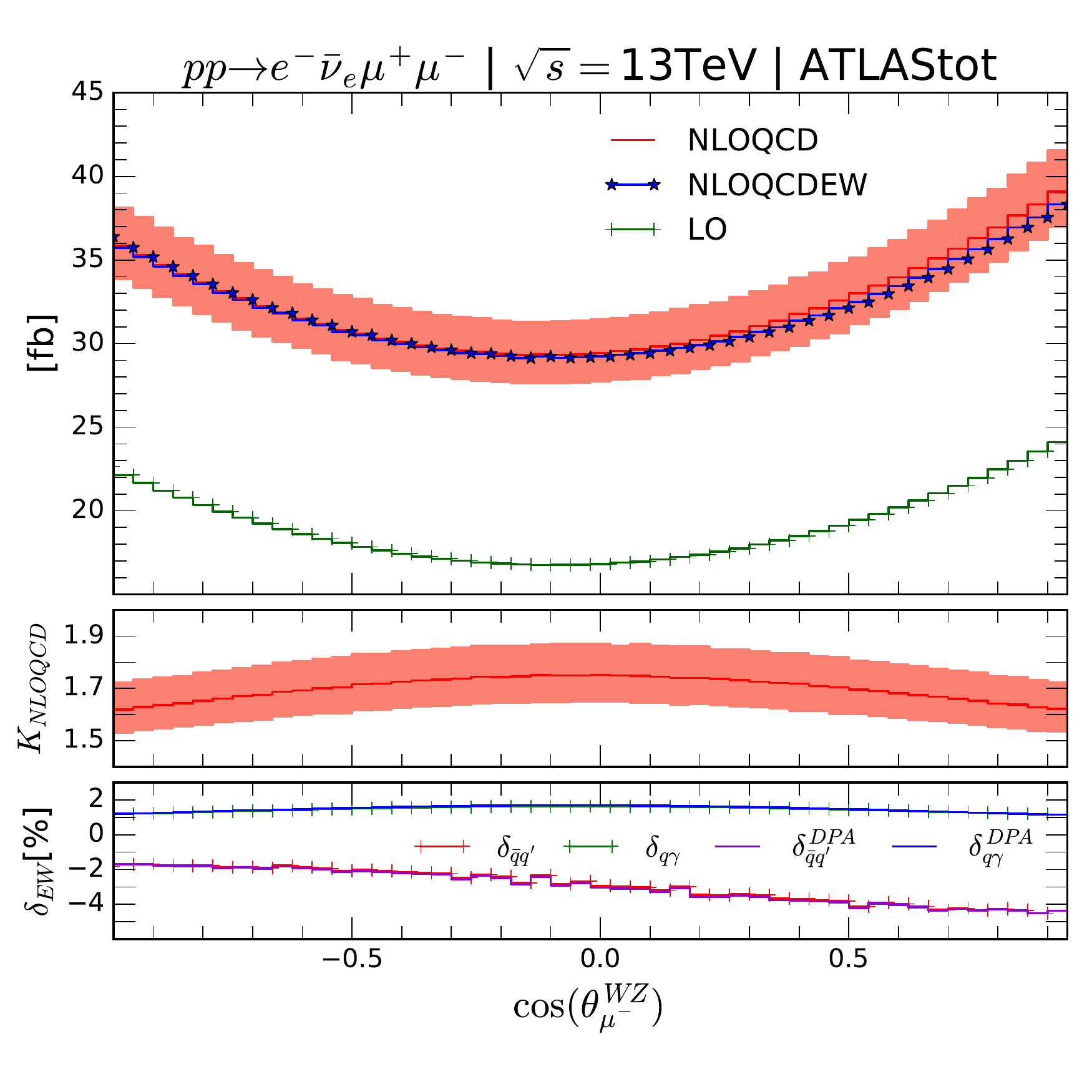}
  \end{tabular}
  \caption{Distributions of the $\cos\theta$ distributions of the
    (anti)electron (left column) and the muon (left column) for the
    process $W^+Z$ (top row) and $W^-Z$ (bottom row). The upper panels
    show the absolute values of the cross sections at LO (in green),
    NLO QCD (red), and NLO QCD+EW (blue). The middle panels display
    the ratio of the NLO QCD cross sections to the corresponding LO
    ones. The bands indicate the total theoretical uncertainty
    calculated as a linear sum of PDF and scale uncertainties at NLO
    QCD. The bottom panels show the NLO EW corrections (see text)
    calculated using DPA relative to the LO (marked with plus
    signs) and DPA LO cross sections.}
  \label{fig:dist_cos_theta_WZ_atlas_total}
\end{figure}

We first present here results for the $\cos\theta$ distributions, from
which the polarization fractions are calculated. They are shown in
\fig{fig:dist_cos_theta_WZ_atlas_total}, where the LO, NLO QCD,
and NLO QCD+EW distributions are separately provided. The bands
indicate the total theoretical uncertainty calculated as a linear sum
of PDF and scale uncertainties at NLO QCD. The $K$ factor defined as
\bea
K_\text{NLOQCD} = \fr{d\sigma_\text{NLOQCD}}{d\sigma_\text{LO}}
\label{eq:K_factor}
\eea
is shown in the middle panels together with the corresponding
uncertainty bands. To quantify the EW corrections, we define, as in
\bib{Baglio:2018rcu}, the following EW corrections
\bea
\delta_{\bar{q}q'} = \fr{d\Delta\sigma^\text{NLOEW}_{\bar{q}q'}}{d\sigma^\text{LO}}, \quad 
\delta_{q\gamma} = \fr{d\Delta\sigma^\text{NLOEW}_{q\gamma}}{d\sigma^\text{LO}},
\label{eq:defs_EW_cor_LO}
\eea %
where the EW corrections to the quark anti-quark annihilation
processes and to the photon quark induced processes are separated. The
reason to show these corrections separately is to see to what extent
they cancel each other. For the case of on-shell $WZ$ production, it
has been shown in \bib{Baglio:2013toa} that this cancellation is
large. In this work, since leptonic decays are included, the QED final
state photon radiation shifts the position of the di-muon invariant
mass, leading to a shift in the photon-radiated contribution to the
$\delta_{\bar{q}q'}$ correction. This shift is negative, making the
$\delta_{\bar{q}q'}$ correction more negative. As a result, we see
that the total EW correction $\delta_\text{EW} = \delta_{\bar{q}q'} +
\delta_{q\gamma}$ is negative, while it is more positive in
\bib{Baglio:2013toa}. Another important difference between this work
and \bib{Baglio:2013toa} is that different photon PDFs are used. This
also changes the $q\gamma$ contribution significantly.

In order to see the effects of the DPA approximation at LO, we replace
the denominators in \eq{eq:defs_EW_cor_LO} by the DPA LO results. This
gives
\bea
\delta_{\bar{q}q'}^\text{DPA} &= \fr{d\Delta\sigma^\text{NLOEW}_{\bar{q}q'}}{d\sigma^\text{LO}_\text{DPA}}, \quad 
\delta_{q\gamma}^\text{DPA} = \fr{d\Delta\sigma^\text{NLOEW}_{q\gamma}}{d\sigma^\text{LO}_\text{DPA}},
\label{eq:defs_EW_cor_DPA}
\eea
which are also shown in the bottom panels in
\fig{fig:dist_cos_theta_WZ_atlas_total} for the sake of
comparison. The EW corrections $\delta_\text{EW}$ are the same when
compared to DPA LO or LO, while in \bib{Baglio:2018rcu} there are some
differences especially at large negative $\cos\theta$ values. The
effect of inclusive cuts is thus here visible.
 
We see that the NLO QCD corrections are large, varying in the range from
$40\%$ to $100\%$ compared to the LO cross section, while the NLO EW
corrections are very small in magnitude, as already
known~\cite{Baglio:2013toa}. However it is important to note that the shape of the angular
distributions is different between the EW corrections and the QCD
corrections, a new feature which has an impact on the polarization
fractions. We see clearly that the QCD corrections are not constant,
but the shape distortion effect is not that large except in the
$\cos\theta_{e^-}$ in the $W^- Z$ channel where the QCD $K$--factor
starts at $K_{NLOQCD}\simeq 1.5$ for large negative $\cos\theta$
values and reaches $K_{NLOQCD}\simeq 1.85$ at large positive values. 
The EW corrections also introduce some visible shape
distortion effects. Comparisons between the $W^+ Z$ and $W^- Z$
channels are valuable as charge asymmetry observables can be
measured. In this context, it is interesting to notice that the QCD
corrections are very similar in the $\cos\theta_{\mu^-}$
distributions, but very different in the $\cos\theta_{e}$
distributions. Remarkably, the opposite behaviors are observed in the
EW corrections, for both $\bar{q}q'$ and $q\gamma$ corrections. The
large charge asymmetry in the QCD corrections to the $\cos\theta_{e}$
distribution is most probably due to the $qg$ induced processes which
first occur at NLO. On the other hand, the large effect observed in
the EW corrections to the $\cos\theta_{\mu^-}$ distribution is due to
the QED final state radiation. This means that the charge asymmetries
in $W$ polarization fractions are more sensitive to the gluon PDF than
in the $Z$ case.

\begin{table}[h!]
  \renewcommand{\arraystretch}{1.3}
\begin{bigcenter}
    \fontsize{8}{8}
\begin{tabular}{|c|c|c|c||c|c|c|}\hline
$\text{Method}$  & $f^{W}_L$ & $f^{W}_0$ & $f^{W}_R$ & $f^Z_L$ & $f^Z_0$ & $f^Z_R$\\
\hline
$\text{DPALO}\,(W^+Z)$ & $0.515$ & $0.153$ & $0.332$ & $0.333$ & $0.144$ & $0.522$\\
\hline
$\text{LO}\,(W^+Z)$ & $0.482(1)^{+1}_{-1}$ & $0.181(1)^{+1}_{-2}$ & $0.337(0.2)^{+1}_{-0.5}$ & $0.306(1)^{+1}_{-1}$ & $0.164(0.4)^{+1}_{-1}$ & $0.529(0.3)^{+0.3}_{-0.2}$\\
\hline
$\text{NLOEW}\,(W^+Z)$ & $0.486$ & $0.180$ & $0.334$ & $0.335$ & $0.169$ & $0.496$\\
\hline
$\text{NLOQCD}\,(W^+Z)$ & $0.471(1)^{+1}_{-1}$ & $0.218(1)^{+3}_{-3}$ & $0.311(1)^{+2}_{-2}$ & $0.338(1)^{+3}_{-3}$ & $0.209(1)^{+4}_{-3}$ & $0.453(1)^{+6}_{-6}$\\
\hline
$\text{NLOQCDEW}\,(W^+Z)$ & $0.473$ & $0.218$ & $0.309$ & $0.355$ & $0.212$ & $0.433$\\
\hline\hline
$\text{DPALO}\,(W^-Z)$ & $0.329$ & $0.158$ & $0.513$ & $0.520$ & $0.150$ & $0.331$\\
\hline
$\text{LO}\,(W^-Z)$ & $0.316(0.4)^{+1}_{-1}$ & $0.181(0.4)^{+1}_{-1}$ & $0.503(0.3)^{+1}_{-1}$ & $0.501(1)^{+1}_{-0.4}$ & $0.168(0.4)^{+1}_{-1}$ & $0.332(0.4)^{+1}_{-1}$\\
\hline
$\text{NLOEW}\,(W^-Z)$ & $0.313$ & $0.181$ & $0.506$ & $0.470$ & $0.172$ & $0.358$\\
\hline
$\text{NLOQCD}\,(W^-Z)$ & $0.344(1)^{+3}_{-2}$ & $0.225(1)^{+3}_{-3}$ & $0.431(1)^{+6}_{-6}$ & $0.478(1)^{+1}_{-2}$ & $0.208(1)^{+3}_{-3}$ & $0.314(1)^{+2}_{-2}$\\
\hline
$\text{NLOQCDEW}\,(W^-Z)$ & $0.342$ & $0.226$ & $0.432$ & $0.459$ & $0.211$ & $0.329$\\
\hline
\end{tabular}
\caption{\small $W$ and $Z$ polarization fractions in the processes
  $pp \to e^+ \nu_e\, \mu^+ \mu^-$ (upper rows) and $pp \to e^-
  \nu_e\, \mu^+ \mu^-$ (lower rows) at DPA LO, LO, NLO EW, NLO QCD,
  and NLO QCD+EW. The PDF uncertainties (in parenthesis) and the scale
  uncertainties are provided for the LO and NLO QCD results, all given
  on the last digit of the central prediction.}
\label{tab:coeff_fL0R_WpmZ}
\end{bigcenter}
\end{table}
\begin{table}[h!]
  \renewcommand{\arraystretch}{1.3}
\begin{center}
\begin{tabular}{|c|c|c|c|c|c|}\hline
$\text{Asymmetry}$  & $\text{DPALO}$ & $\text{LO}$ & $\text{NLOEW}$ & $\text{NLOQCD}$ & $\text{NLOQCDEW}$\\
\hline
$\mathcal{A}^W_{LR}$ & $-27.0\%$ & $-12.7\%$ & $-11.9\%$ & $+29.6\%$ & $+29.1\%$ \\
\hline
$\mathcal{A}^Z_{LR}$ & $0\%$ & $+13.8\%$ & $+17.9\%$ & $-17.6\%$ & $-25.0\%$ \\
\hline
\end{tabular}
    \caption{\small Left-right charge asymmetries.}
    \label{tab:coeff_asym_LR}
\end{center}
\end{table}
From the above $\cos\theta$ distributions, the polarization fractions
are calculated. This result is presented in \tab{tab:coeff_fL0R_WpmZ},
where PDF and scale uncertainties associated with the LO and NLO QCD
predictions are also calculated. To quantify the aforementioned
higher-order effects on charge asymmetry observables, we define here
two observables,
\bea
\mathcal{A}^V_{LR} = \fr{q^V_{W^+ Z}-q^V_{W^- Z}}{q^V_{W^+ Z}+q^V_{W^- Z}}, \quad 
\mathcal{B}^V_{0} = \fr{p^V_{W^+ Z}-p^V_{W^- Z}}{p^V_{W^+ Z}+p^V_{W^- Z}},
\eea
where $q^V = |f^V_L - f^V_R|$ and $p^V = |f^V_0|$. Note
that, absolute values are needed because, in general, the fractions
can get negative as shown in \bib{Baglio:2018rcu} for
the case of the fiducial distributions in the Collins-Soper coordinate
system. Furthermore, since the three fractions sum to unity,
only two parameters are independent. From \tab{tab:coeff_fL0R_WpmZ} we
see that $\mathcal{B}^V_{0}$ is not interesting as this asymmetry is
very small, namely $\mathcal{B}^W_{0} \approx -2\%$ and
$\mathcal{B}^Z_{0} \approx +0.2\%$ at NLO QCD+EW accuracy. However,
the left-right charge asymmetries $\mathcal{A}^{V}_{LR}$ are
much larger, being $\mathcal{A}^W_{LR} \approx +29\%$ and
$\mathcal{A}^Z_{LR} \approx -25\%$ at NLO QCD+EW accuracy. Results at
DPA LO, LO, NLO EW and NLO QCD levels are provided in
\tab{tab:coeff_asym_LR}, showing that these observables are very
sensitive to off-shell and higher-order effects 
as the left-right asymmetries are negligible in the DPA limit, 
being  numerically at the per mill level for the $W$ case and even
smaller for the $Z$ case. Consistently with the above
observations on the distributions, we see that the $W$
asymmetry is more sensitive to the QCD corrections, while the $Z$
asymmetry is more sensitive to the EW corrections.
Note that the theory uncertainties cancel in the ratios 
defining the left-right asymmetries and are then expected to be negligible. 
We will not discuss them further.

\begin{table}[h!]
  \renewcommand{\arraystretch}{1.3}
\begin{center}
\begin{tabular}{|c|c|c|c||c|c|c|}\hline
$\text{Cut}$  & $f^{W}_L$ & $f^{W}_0$ & $f^{W}_R$ & $f^Z_L$ & $f^Z_0$ & $f^Z_R$\\
\hline
$\text{CUT-1}\,(W^+Z)$ & $0.475$ & $0.218$ & $0.307$ & $0.360$ & $0.217$ & $0.424$\\
$\text{CUT-2}\,(W^+Z)$ & $0.475$ & $0.217$ & $0.308$ & $0.359$ & $0.214$ & $0.427$\\
$\text{CUT-3}\,(W^+Z)$ & $0.475$ & $0.217$ & $0.308$ & $0.357$ & $0.213$ & $0.429$\\
$\text{CUT-4}\,(W^+Z)$ & $0.474$ & $0.217$ & $0.309$ & $0.356$ & $0.213$ & $0.431$\\
$\text{CUT-5}\,(W^+Z)$ & $0.473$ & $0.218$ & $0.309$ & $0.355$ & $0.212$ & $0.433$\\
$\text{CUT-6}\,(W^+Z)$ & $0.473$ & $0.218$ & $0.310$ & $0.353$ & $0.212$ & $0.435$\\
\hline\hline
$\text{CUT-1}\,(W^-Z)$ & $0.341$ & $0.227$ & $0.432$ & $0.455$ & $0.215$ & $0.329$\\
$\text{CUT-2}\,(W^-Z)$ & $0.341$ & $0.226$ & $0.433$ & $0.457$ & $0.213$ & $0.330$\\
$\text{CUT-3}\,(W^-Z)$ & $0.341$ & $0.226$ & $0.433$ & $0.458$ & $0.212$ & $0.330$\\
$\text{CUT-4}\,(W^-Z)$ & $0.342$ & $0.226$ & $0.433$ & $0.459$ & $0.212$ & $0.329$\\
$\text{CUT-5}\,(W^-Z)$ & $0.342$ & $0.226$ & $0.432$ & $0.459$ & $0.211$ & $0.329$\\
$\text{CUT-6}\,(W^-Z)$ & $0.343$ & $0.226$ & $0.432$ & $0.460$ & $0.211$ & $0.329$\\
\hline
\end{tabular}
    \caption{\small NLO QCD+EW $W$ and $Z$ polarization fractions at different cuts (see text).}
    \label{tab:coeff_fL0R_WpmZ_cuts}
\end{center}
\end{table}
We close this section by presenting in \tab{tab:coeff_fL0R_WpmZ_cuts}
the NLO QCD+EW results of the fractions for different invariant mass
windows of the muon pair. These cuts are named CUT-i, $i=1,\ldots,6$,
corresponding respectively to 
(86 GeV, 96 GeV), (81 GeV, 101 GeV), (76 GeV, 106 GeV), (71 GeV, 111 GeV), (66 GeV, 116 GeV), and (60 GeV, 120 GeV). 
Note that CUT-5 is
the default cut defined in \eq{eq:cut_default}, while CUT-6 is used by
CMS~\cite{Khachatryan:2016tgp} to define their total phase space. As
expected, we see that the $W$ fractions are almost unchanged while the
$Z$ fractions vary more visibly. For the $Z$ fractions, the different
behaviors between the $W^+ Z$ and $W^- Z$ channels are
interesting. While the $f_R^Z$ in the former case varies most
strongly, it is almost unchanged in the latter. This unexpected
constant behavior is due to the opposite behaviors in the other
fractions.
 
\subsection{Angular coefficients}
\label{sect:polar_observables_wp}
We now turn to the angular coefficients. Results for the
(anti)electron are presented in \tab{tab:coeff_Ai_e} and for the muon
in \tab{tab:coeff_Ai_mu} at LO, NLO QCD, NLO EW, and NLO
QCD+EW levels. 
\begin{table}[ht!]
 \renewcommand{\arraystretch}{1.3}
\begin{bigcenter}
\setlength\tabcolsep{0.03cm}
\fontsize{7.0}{7.0}
\begin{tabular}{|c|c|c|c|c|c|c|c|c|}\hline
$\text{Method}$  & $A_0$ & $A_1$  & $A_2$ & $A_3$ & $A_4$ & $A_5$ & $A_6$ & $A_7$\\
\hline
{\fontsize{6.0}{6.0}$\text{LO}\,(W^+Z)$} & $0.363(1)^{+3}_{-3}$ & $-0.049(1)^{+1}_{-2}$ & $-0.232(1)^{+4}_{-4}$ & $-0.102(2)^{+1}_{-2}$ & $-0.289(1)^{+1}_{-1}$ & $-0.0002(1)^{+1}_{-1}$ & $-0.003(0.3)^{+0.1}_{-0.01}$ & $-0.011(0.5)^{+0.2}_{-0.03}$\\
\hline
{\fontsize{6.0}{6.0}$\text{NLOEW}\,(W^+Z)$} & $0.359$ & $-0.052$ & $-0.215$ & $-0.091$ & $-0.304$ & $-0.001$ & $-0.005$ & $-0.007$\\
\hline
{\fontsize{6.0}{6.0}$\text{NLOQCD}\,(W^+Z)$} & $0.436(1)^{+6}_{-5}$ & $-0.120(1)^{+6}_{-6}$ & $-0.233(1)^{+3}_{-2}$ & $-0.021(2)^{+7}_{-7}$ & $-0.319(2)^{+3}_{-3}$ & $0.0003(5)^{+2}_{-0.4}$ & $-0.001(1)^{+0.01}_{-0.2}$ & $-0.001(0.4)^{+1}_{-1}$\\
\hline
{\fontsize{6.0}{6.0}$\text{NLOQCDEW}\,(W^+Z)$} & $0.435$ & $-0.122$ & $-0.222$ & $-0.014$ & $-0.328$ & $-0.0002$ & $-0.002$ & $0.001$\\
\hline\hline
{\fontsize{6.0}{6.0}$\text{LO}\,(W^-Z)$} & $0.362(1)^{+3}_{-3}$ & $-0.067(1)^{+2}_{-2}$ & $-0.253(1)^{+4}_{-3}$ & $0.118(2)^{+4}_{-3}$ & $-0.374(1)^{+0.2}_{-0.2}$ & $0.002(0.2)^{+0}_{-0.03}$ & $0.002(0.4)^{+0}_{-0.1}$ & $0.007(0.4)^{+0.2}_{-0.02}$\\
\hline
{\fontsize{6.0}{6.0}$\text{NLOEW}\,(W^-Z)$} & $0.362$ & $-0.072$ & $-0.232$ & $0.127$ & $-0.388$ & $0.003$ & $0.004$ & $0.0005$\\
\hline
{\fontsize{6.0}{6.0}$\text{NLOQCD}\,(W^-Z)$} & $0.451(1)^{+7}_{-7}$ & $-0.131(1)^{+5}_{-5}$ & $-0.233(2)^{+4}_{-4}$ & $0.032(2)^{+7}_{-8}$ & $-0.174(4)^{+17}_{-16}$ & $0.001(1)^{+0.3}_{-0.1}$ & $0.001(0.4)^{+0.2}_{-0.1}$ & $-0.005(0.3)^{+1}_{-1}$\\
\hline
{\fontsize{6.0}{6.0}$\text{NLOQCDEW}\,(W^-Z)$} & $0.451$ & $-0.135$ & $-0.219$ & $0.037$ & $-0.180$ & $0.002$ & $0.002$ & $-0.009$\\
\hline
\end{tabular}
\caption{\small Angular coefficients of the $e^{+}$ (upper rows) and
  $e^-$ (lower rows) distributions for the final states $e^+ \nu_e\,
  \mu^+ \mu^-$ and $e^- \bar{\nu}_e\, \mu^+ \mu^-$, respectively, at
  LO, NLO EW, NLO QCD, and NLO QCD+EW. The PDF uncertainties (in
  parenthesis) and the scale uncertainties are provided for the LO and
  NLO QCD results, all given on the last digit of the central
  prediction.}
\label{tab:coeff_Ai_e}
\end{bigcenter}
\end{table}
\begin{table}[ht!]
 \renewcommand{\arraystretch}{1.3}
\begin{bigcenter}
\setlength\tabcolsep{0.03cm}
\fontsize{7.0}{7.0}
\begin{tabular}{|c|c|c|c|c|c|c|c|c|}\hline
$\text{Method}$  & $A_0$ & $A_1$  & $A_2$ & $A_3$ & $A_4$ & $A_5$ & $A_6$ & $A_7$\\
\hline
{\fontsize{6.0}{6.0}$\text{LO}\,(W^+Z)$} & $0.329(1)^{+2}_{-3}$ & $-0.034(1)^{+2}_{-2}$ & $-0.145(1)^{+3}_{-2}$ & $0.032(1)^{+1}_{-1}$ & $-0.096(0.3)^{+0.4}_{-0.3}$ & $0.000004(209)^{+87}_{-83}$ & $-0.011(0.2)^{+0.1}_{-0.1}$ & $0.013(0.2)^{+0.05}_{-0.2}$\\
\hline
{\fontsize{6.0}{6.0}$\text{NLOEW}\,(W^+Z)$} & $0.338$ & $-0.038$ & $-0.136$ & $0.041$ & $-0.069$ & $-0.0005$ & $-0.012$ & $0.013$\\
\hline
{\fontsize{6.0}{6.0}$\text{NLOQCD}\,(W^+Z)$} & $0.418(1)^{+7}_{-7}$ & $-0.096(1)^{+5}_{-6}$ & $-0.144(1)^{+2}_{-2}$ & $0.010(1)^{+2}_{-2}$ & $-0.049(1)^{+4}_{-4}$ & $0.0002(10)^{+2}_{-0.04}$ & $-0.008(0.3)^{+0.2}_{-0.2}$ & $0.009(0.2)^{+0.2}_{-0.3}$\\
\hline
{\fontsize{6.0}{6.0}$\text{NLOQCDEW}\,(W^+Z)$} & $0.424$ & $-0.100$ & $-0.139$ & $0.015$ & $-0.033$ & $-0.0002$ & $-0.009$ & $0.009$\\
\hline\hline
{\fontsize{6.0}{6.0}$\text{LO}\,(W^-Z)$} & $0.335(1)^{+3}_{-3}$ & $0.013(1)^{+2}_{-2}$ & $-0.153(1)^{+2}_{-2}$ & $0.012(0.4)^{+0.2}_{-0.1}$ & $0.072(0.3)^{+0.2}_{-0.2}$ & $0.004(0.2)^{+0.1}_{-0.04}$ & $0.009(0.2)^{+0.1}_{-0.1}$ & $0.014(0.2)^{+0.1}_{-0.2}$\\
\hline
{\fontsize{6.0}{6.0}$\text{NLOEW}\,(W^-Z)$} & $0.344$ & $0.008$ & $-0.146$ & $0.014$ & $0.048$ & $0.004$ & $0.011$ & $0.014$\\
\hline
{\fontsize{6.0}{6.0}$\text{NLOQCD}\,(W^-Z)$} & $0.417(1)^{+6}_{-6}$ & $-0.071(1)^{+7}_{-7}$ & $-0.164(1)^{+2}_{-1}$ & $0.001(0.5)^{+1}_{-1}$ & $0.070(0.5)^{+0.3}_{-0.3}$ & $0.002(0.5)^{+0.2}_{-0.1}$ & $0.006(1)^{+0.2}_{-0.2}$ & $0.009(1)^{+0.3}_{-0.3}$\\
\hline
{\fontsize{6.0}{6.0}$\text{NLOQCDEW}\,(W^-Z)$} & $0.423$ & $-0.075$ & $-0.160$ & $0.002$ & $0.056$ & $0.003$ & $0.007$ & $0.009$\\
\hline
\end{tabular}
\caption{\small Same as \tab{tab:coeff_Ai_e} but for the $\mu^-$ coefficients.}
\label{tab:coeff_Ai_mu}
\end{bigcenter}
\end{table} 

The first thing to notice is that the values of the P-odd coefficients
$A_{5-7}$ are very small, but non-vanishing. In order to see that they
are indeed statistically non-zero, we show the DPA LO and LO results
together with statistical errors in \tab{tab:coeff_Ai_DPA_LO_e} for
the (anti)electron case. For completeness, similar results for the
muon case are also provided in \tab{tab:coeff_Ai_DPA_LO_mu}. We see
that those coefficients are zero within the statistical error in the
DPA. However, at LO, when full off-shell effects are taken into
account, they are all non-zero, except $A_5^{e^+}$ and $A_{5,
  W^+Z}^{\mu^-}$ where the results are very small. It has been shown
in \cite{Aguilar-Saavedra:2015yza,Baglio:2018rcu} that, in the DPA,
$A_{5-7}$ are proportional to the imaginary part of the spin-density
matrices. This means that in the zero-width limit (i.e. $\Gamma_V \to
0$) they are vanishing, explaining why they are so small in the
DPA. Note that, a finite width induces an off-shell effect, but the
full off-shell effects at LO include additionally new Feynman diagrams
such as virtual photon and single-resonant contributions. This
explains why the values of the P-odd coefficients are significantly
larger at LO than at DPA LO. Results in \tab{tab:coeff_Ai_e} and
\tab{tab:coeff_Ai_mu} show that they remain very small at NLO QCD+EW
accuracy. Measuring them in experiments is therefore very challenging (see \cite{Frederix:2014cba} 
for a study on $W$-jet production at the LHC). 
\begin{table}[h!]
 \renewcommand{\arraystretch}{1.3}
\begin{center}
\setlength\tabcolsep{0.1cm}
\fontsize{9.0}{9.0}
\begin{tabular}{|c|c|c|c|c|c|c|c|c|}\hline
$\text{Method}$  & $A_0$ & $A_1$  & $A_2$ & $A_3$ & $A_4$ & $A_5$ & $A_6$ & $A_7$\\
\hline
{\fontsize{9.0}{9.0}$\text{DPALO}\,(W^+Z)$} & $0.306$ & $-0.030$ & $-0.286$ & $0.092$ & $-0.368$ & $-0.00001[40]$ & $-0.0001[2]$ & $0.00005[19]$\\
\hline
{\fontsize{9.0}{9.0}$\text{LO}\,(W^+Z)$} & $0.363$ & $-0.049$ & $-0.232$ & $-0.102$ & $-0.289$ & $-0.0002[5]$ & $-0.0034[4]$ & $-0.011$\\
\hline\hline
{\fontsize{9.0}{9.0}$\text{DPALO}\,(W^-Z)$} & $0.317$ & $-0.0023[2]$ & $-0.295$ & $-0.082$ & $-0.367$ & $0.00001[38]$ & $-0.0001[2]$ & $0.0001[2]$\\
\hline
{\fontsize{9.0}{9.0}$\text{LO}\,(W^-Z)$} & $0.362$ & $-0.066$ & $-0.254$ & $0.118$ & $-0.374$ & $0.0024[4]$ & $0.0024[3]$ & $0.0067[2]$\\
\hline
\end{tabular}
\caption{\small Angular coefficients of the $e^{+}$ (upper rows) and
  $e^-$ (lower rows) distributions for the final states $e^+ \nu_e\,
  \mu^+ \mu^-$ and $e^- \bar{\nu}_e\, \mu^+ \mu^-$, respectively, at
  DPA LO and LO. The numbers in square brackets represent the
  statistical error, when it is significant.}
\label{tab:coeff_Ai_DPA_LO_e}
\end{center}
\end{table} 
\begin{table}[h!]
 \renewcommand{\arraystretch}{1.3}
\begin{center}
\setlength\tabcolsep{0.1cm}
\fontsize{9.0}{9.0}
\begin{tabular}{|c|c|c|c|c|c|c|c|c|}\hline
$\text{Method}$  & $A_0$ & $A_1$  & $A_2$ & $A_3$ & $A_4$ & $A_5$ & $A_6$ & $A_7$\\
\hline
{\fontsize{9.0}{9.0}$\text{DPALO}\,(W^+Z)$} & $0.289$ & $0.037$ & $-0.183$ & $-0.026$ & $-0.081$ & $-0.0001[2]$ & $-0.0001[1]$ & $0.00001[17]$\\
\hline
{\fontsize{9.0}{9.0}$\text{LO}\,(W^+Z)$} & $0.329$ & $-0.034$ & $-0.145$ & $0.032$ & $-0.096$ & $0.0001[4]$ & $-0.011$ & $0.013$\\
\hline\hline
{\fontsize{9.0}{9.0}$\text{DPALO}\,(W^-Z)$} & $0.299$ & $0.026$ & $-0.188$ & $-0.0085[1]$ & $0.081$ & $-0.00003[19]$ & $-0.0001[2]$ & $0.00002[14]$\\
\hline
{\fontsize{9.0}{9.0}$\text{LO}\,(W^-Z)$} & $0.336$ & $0.013$ & $-0.154$ & $0.012$ & $0.072$ & $0.0041[8]$ & $0.0091[4]$ & $0.014$\\
\hline
\end{tabular}
\caption{\small Same as \tab{tab:coeff_Ai_DPA_LO_e} but for the
  $\mu^-$ distributions.}
\label{tab:coeff_Ai_DPA_LO_mu}
\end{center}
\end{table} 

We now focus on the P-even coefficients $A_{1-4}$. Results at the NLO
QCD+EW accuracy in \tab{tab:coeff_Ai_e} and \tab{tab:coeff_Ai_mu} show
that they are large, hence should be measureable. EW corrections are
significant for $A^{\mu^-}_3$ and $A^{\mu^-}_4$ due to the radiative
corrections to the $Z \to \mu^+ \mu^-$ decay, as found in
\bib{Baglio:2018rcu} using more exclusive cuts.

Similarly to the previous section, we can define various charge
asymmetries as
\bea
\mathcal{C}^V_{i} = \fr{r^V_{i,W^+ Z}-r^V_{i,W^- Z}}{r^V_{i,W^+ Z}+r^V_{i,W^- Z}}, \;\;\;
r^W_{i} = |A^e_i|,\;\; 
r^Z_{i} = |A^{\mu^-}_i|,\;\; i=1,\ldots,7.
\eea
Because of the relations between $f_{L,R,0}$ and $A_{0,4}$ 
presented in Eq.(\ref{eq:relations_fL0R_Ai}), the asymmetries
$\mathcal{C}^V_{0}$ and $\mathcal{C}^V_{4}$ are identical to
$\mathcal{B}^V_{0}$ and $\mathcal{A}^V_{LR}$, respectively. Looking at
\tab{tab:coeff_Ai_e} and \tab{tab:coeff_Ai_mu} we see that
$\mathcal{C}^V_{3}$ and $\mathcal{C}^V_{4}$ are the most
interesting. Note that we ignore $\mathcal{C}^V_{5-7}$ in this
discussion because they are very difficult to measure. Similar to
\tab{tab:coeff_asym_LR}, values for $\mathcal{C}^V_{3}$ in different
approximations are provided in \tab{tab:coeff_asym_A3}. Compared to
$\mathcal{A}^V_{LR}$, this asymmetry is much larger. Another
interesting difference is the value at DPA LO. While it is almost
vanishing for $\mathcal{A}^V_{LR}$, it is large for
$\mathcal{C}^V_{3}$, reaching $+51\%$ for the $Z$ case, suggesting
very different origins. This is indeed the case because $A_3$ is
related to the off-diagonal entries of the spin-density matrices at
DPA LO, while $A_4$ comes from the diagonal ones
\cite{Baglio:2018rcu}. This is also why the value of $A_3$  is much
smaller than that of $A_4$. All in all, we see that the value of
$\mathcal{C}^V_{3}$ is much larger hence could be an additional probe
of the spin structure of the gauge bosons, but this observable is more
difficult to measure than $\mathcal{A}^V_{LR}$ because of difficulties
in measuring $A_3$.

\begin{table}[h!]
  \renewcommand{\arraystretch}{1.3}
\begin{center}
\begin{tabular}{|c|c|c|c|c|c|}\hline
$\text{Asymmetry}$  & $\text{DPALO}$ & $\text{LO}$ & $\text{NLOEW}$ & $\text{NLOQCD}$ & $\text{NLOQCDEW}$\\
\hline
$\mathcal{C}^W_{3}$ & $+0.057$ & $-0.073$ & $-0.165$ & $-0.208$ & $-0.451$ \\
\hline
$\mathcal{C}^Z_{3}$ & $+0.507$ & $+0.455$ & $+0.491$ & $+0.818$ & $+0.765$ \\
\hline
\end{tabular}
    \caption{\small Charge asymmetries in the coefficient $A_3$.}
    \label{tab:coeff_asym_A3}
\end{center}
\end{table}

\section{Conclusions}
\label{sect:conclusion}
In this paper, we have presented results for polarization fractions
and angular coefficients up to the NLO QCD+EW accuracy in the process
$p p \to e^\pm \nu_e \mu^+ \mu^-$ at the 13 TeV LHC.
We have used by default the total phase space adopted by ATLAS in
their latest 13 TeV analysis of the $W$ and $Z$ polarization
observables, measured in the modified helicity coordinate system where
the $z$-axis is defined as the gauge boson flight direction 
in the $WZ$ center-of-mass frame. Our LO
and NLO QCD results include full off-shell effects while the EW corrections
have been calculated in the double-pole approximation.

We have defined left-right charge asymmetries $\mathcal{A}_{LR}^{W/Z}$ out of
the polarization fractions that are very sensitive to higher-order
effects, in particular $\mathcal{A}_{LR}^{W}$ is sensitive to the QCD
corrections while $\mathcal{A}_{LR}^{Z}$ is sensitive to the EW
corrections. This sensitivity can be related to the shape of the
higher-order corrections to the angular distributions underlying the
calculations of the polarization fractions. These asymmetries are
found to be large at NLO QCD+EW order, of the order of $+29\%$ for the
$W$ boson and $-25\%$ for the $Z$ boson, that should be measurable at
the LHC. 

We have also defined similar charge asymmetries for the angular
coefficients themselves and we have found that the asymmetries related
to the P-even coefficient $A_3$ can also be very large, of the
order of $-45\%$ for the $W$ boson and $+78\%$ for the $Z$ boson, and
could be additional probes of the polarization structure of the underlying dynamics, 
even if they are more difficult to measure than the left-right charge 
asymmetries. The EW corrections are also found to be sizable in these $A_3$ 
charge asymmetries. 

This work is a step towards a study of the polarization observables
that is as close as possible to the current experimental setup in
ATLAS.

\acknowledgments

This work is funded by the Vietnam National Foundation for Science and
Technology Development (NAFOSTED) under grant number
103.01-2017.78. We are grateful to Emmanuel Sauvan for fruitful
discussions.



\begin{thebibliography}{10}

\bibitem{Aaboud:2019gxl}
{\scshape ATLAS} collaboration, M.~Aaboud et~al., \emph{{Measurement of
  $W^{\pm}Z$ production cross sections and gauge boson polarisation in $pp$
  collisions at $\sqrt{s} = 13$ TeV with the ATLAS detector}},
  \href{https://doi.org/10.1140/epjc/s10052-019-7027-6}{\emph{Eur. Phys. J.}
  {\bfseries C79} (2019) 535}
  [\href{https://arxiv.org/abs/1902.05759}{{\ttfamily 1902.05759}}].

\bibitem{Bilchak:1984gv}
C.~L. Bilchak, R.~W. Brown and J.~D. Stroughair, \emph{{$W^\pm_{}$ and $Z^0_{}$
  Polarization in Pair Production: Dominant Helicities}},
  \href{https://doi.org/10.1103/PhysRevD.29.375}{\emph{Phys. Rev.} {\bfseries
  D29} (1984) 375}.

\bibitem{Willenbrock:1987xz}
S.~S.~D. Willenbrock, \emph{{Pair Production of $W$ and $Z$ Bosons and the
  Goldstone Boson Equivalence Theorem}},
  \href{https://doi.org/10.1016/S0003-4916(88)80016-2}{\emph{Annals Phys.}
  {\bfseries 186} (1988) 15}.

\bibitem{Stirling:2012zt}
W.~J. Stirling and E.~Vryonidou, \emph{{Electroweak gauge boson polarisation at
  the LHC}}, \href{https://doi.org/10.1007/JHEP07(2012)124}{\emph{JHEP}
  {\bfseries 07} (2012) 124} [\href{https://arxiv.org/abs/1204.6427}{{\ttfamily
  1204.6427}}].

\bibitem{Baglio:2018rcu}
J.~Baglio and L.~D. Ninh, \emph{{Fiducial polarization observables in hadronic
  WZ production: A next-to-leading order QCD+EW study}},
  \href{https://doi.org/10.1007/JHEP04(2019)065}{\emph{JHEP} {\bfseries 04}
  (2019) 065} [\href{https://arxiv.org/abs/1810.11034}{{\ttfamily
  1810.11034}}].

\bibitem{Biedermann:2017oae}
B.~Biedermann, A.~Denner and L.~Hofer, \emph{{Next-to-leading-order electroweak
  corrections to the production of three charged leptons plus missing energy at
  the LHC}}, \href{https://doi.org/10.1007/JHEP10(2017)043}{\emph{JHEP}
  {\bfseries 10} (2017) 043}
  [\href{https://arxiv.org/abs/1708.06938}{{\ttfamily 1708.06938}}].

\bibitem{Bierweiler:2013dja}
A.~Bierweiler, T.~Kasprzik and J.~H. K{\"u}hn, \emph{{Vector-boson pair
  production at the LHC to $\mathcal{O}(\alpha^3)$ accuracy}},
  \href{https://doi.org/10.1007/JHEP12(2013)071}{\emph{JHEP} {\bfseries 12}
  (2013) 071} [\href{https://arxiv.org/abs/1305.5402}{{\ttfamily 1305.5402}}].

\bibitem{Baglio:2013toa}
J.~Baglio, L.~D. Ninh and M.~M. Weber, \emph{{Massive gauge boson pair
  production at the LHC: a next-to-leading order story}},
  \href{https://doi.org/10.1103/PhysRevD.94.099902,
  10.1103/PhysRevD.88.113005}{\emph{Phys. Rev.} {\bfseries D88} (2013) 113005}
  [\href{https://arxiv.org/abs/1307.4331}{{\ttfamily 1307.4331}}].

\bibitem{Ohnemus:1991gb}
J.~Ohnemus, \emph{{An order $\alpha_s$ calculation of hadronic $W^\pm Z$
  production}}, \href{https://doi.org/10.1103/PhysRevD.44.3477}{\emph{Phys.
  Rev.} {\bfseries D44} (1991) 3477}.

\bibitem{Frixione:1992pj}
S.~Frixione, P.~Nason and G.~Ridolfi, \emph{{Strong corrections to W Z
  production at hadron colliders}},
  \href{https://doi.org/10.1016/0550-3213(92)90668-2}{\emph{Nucl. Phys.}
  {\bfseries B383} (1992) 3}.

\bibitem{Grazzini:2016swo}
M.~Grazzini, S.~Kallweit, D.~Rathlev and M.~Wiesemann, \emph{{$W^{\pm}Z$
  production at hadron colliders in NNLO QCD}},
  \href{https://doi.org/10.1016/j.physletb.2016.08.017}{\emph{Phys. Lett.}
  {\bfseries B761} (2016) 179}
  [\href{https://arxiv.org/abs/1604.08576}{{\ttfamily 1604.08576}}].

\bibitem{Grazzini:2017ckn}
M.~Grazzini, S.~Kallweit, D.~Rathlev and M.~Wiesemann, \emph{{$W^\pm Z$
  production at the LHC: fiducial cross sections and distributions in NNLO
  QCD}}, \href{https://doi.org/10.1007/JHEP05(2017)139}{\emph{JHEP} {\bfseries
  05} (2017) 139} [\href{https://arxiv.org/abs/1703.09065}{{\ttfamily
  1703.09065}}].

\bibitem{Collins:1977iv}
J.~C. Collins and D.~E. Soper, \emph{{Angular Distribution of Dileptons in
  High-Energy Hadron Collisions}},
  \href{https://doi.org/10.1103/PhysRevD.16.2219}{\emph{Phys. Rev.} {\bfseries
  D16} (1977) 2219}.

\bibitem{Bern:2011ie}
Z.~Bern et~al., \emph{{Left-Handed W Bosons at the LHC}},
  \href{https://doi.org/10.1103/PhysRevD.84.034008}{\emph{Phys. Rev.}
  {\bfseries D84} (2011) 034008}
  [\href{https://arxiv.org/abs/1103.5445}{{\ttfamily 1103.5445}}].

\bibitem{Arnold:2008rz}
K.~Arnold et~al., \emph{{VBFNLO: A Parton level Monte Carlo for processes with
  electroweak bosons}},
  \href{https://doi.org/10.1016/j.cpc.2009.03.006}{\emph{Comput. Phys. Commun.}
  {\bfseries 180} (2009) 1661}
  [\href{https://arxiv.org/abs/0811.4559}{{\ttfamily 0811.4559}}].

\bibitem{Baglio:2014uba}
J.~Baglio et~al., \emph{{Release Note - VBFNLO 2.7.0}},
  \href{https://arxiv.org/abs/1404.3940}{{\ttfamily 1404.3940}}.

\bibitem{Chaichian:1981va}
M.~Chaichian, M.~Hayashi and K.~Yamagishi, \emph{{Angular Distributions of High
  Mass Dileptons With Finite Transverse Momentum in High-energy Hadronic
  Collisions}}, \href{https://doi.org/10.1103/PhysRevD.25.130,
  10.1103/PhysRevD.26.2534}{\emph{Phys. Rev.} {\bfseries D25} (1982) 130}.

\bibitem{Mirkes:1992hu}
E.~Mirkes, \emph{{Angular decay distribution of leptons from W bosons at NLO in
  hadronic collisions}},
  \href{https://doi.org/10.1016/0550-3213(92)90046-E}{\emph{Nucl. Phys.}
  {\bfseries B387} (1992) 3}.

\bibitem{Hagiwara:1984hi}
K.~Hagiwara, K.-i. Hikasa and N.~Kai, \emph{{Parity Odd Asymmetries in $W$ Jet
  Events at Hadron Colliders}},
  \href{https://doi.org/10.1103/PhysRevLett.52.1076}{\emph{Phys. Rev. Lett.}
  {\bfseries 52} (1984) 1076}.

\bibitem{Frederix:2014cba}
R.~Frederix, K.~Hagiwara, T.~Yamada and H.~Yokoya, \emph{{T-odd Asymmetry in
  W+jet Events at the LHC}},
  \href{https://doi.org/10.1103/PhysRevLett.113.152001}{\emph{Phys. Rev. Lett.}
  {\bfseries 113} (2014) 152001}
  [\href{https://arxiv.org/abs/1407.1016}{{\ttfamily 1407.1016}}].

\bibitem{Aguilar-Saavedra:2015yza}
J.~A. Aguilar-Saavedra and J.~Bernabeu, \emph{{Breaking down the entire W boson
  spin observables from its decay}},
  \href{https://doi.org/10.1103/PhysRevD.93.011301}{\emph{Phys. Rev.}
  {\bfseries D93} (2016) 011301}
  [\href{https://arxiv.org/abs/1508.04592}{{\ttfamily 1508.04592}}].

\bibitem{Aguilar-Saavedra:2017zkn}
J.~A. Aguilar-Saavedra, J.~{Bernab\'eu}, V.~A. Mitsou and A.~Segarra,
  \emph{{The Z boson spin observables as messengers of new physics}},
  \href{https://doi.org/10.1140/epjc/s10052-017-4795-8}{\emph{Eur. Phys. J.}
  {\bfseries C77} (2017) 234}
  [\href{https://arxiv.org/abs/1701.03115}{{\ttfamily 1701.03115}}].

\bibitem{Manohar:2016nzj}
A.~Manohar, P.~Nason, G.~P. Salam and G.~Zanderighi, \emph{{How bright is the
  proton? A precise determination of the photon parton distribution function}},
  \href{https://doi.org/10.1103/PhysRevLett.117.242002}{\emph{Phys. Rev. Lett.}
  {\bfseries 117} (2016) 242002}
  [\href{https://arxiv.org/abs/1607.04266}{{\ttfamily 1607.04266}}].

\bibitem{Manohar:2017eqh}
A.~V. Manohar, P.~Nason, G.~P. Salam and G.~Zanderighi, \emph{{The Photon
  Content of the Proton}},
  \href{https://doi.org/10.1007/JHEP12(2017)046}{\emph{JHEP} {\bfseries 12}
  (2017) 046} [\href{https://arxiv.org/abs/1708.01256}{{\ttfamily
  1708.01256}}].

\bibitem{Butterworth:2015oua}
J.~Butterworth et~al., \emph{{PDF4LHC recommendations for LHC Run II}},
  \href{https://doi.org/10.1088/0954-3899/43/2/023001}{\emph{J. Phys.}
  {\bfseries G43} (2016) 023001}
  [\href{https://arxiv.org/abs/1510.03865}{{\ttfamily 1510.03865}}].

\bibitem{Dulat:2015mca}
S.~Dulat, T.-J. Hou, J.~Gao, M.~Guzzi, J.~Huston, P.~Nadolsky et~al.,
  \emph{{New parton distribution functions from a global analysis of quantum
  chromodynamics}},
  \href{https://doi.org/10.1103/PhysRevD.93.033006}{\emph{Phys. Rev.}
  {\bfseries D93} (2016) 033006}
  [\href{https://arxiv.org/abs/1506.07443}{{\ttfamily 1506.07443}}].

\bibitem{Harland-Lang:2014zoa}
L.~A. Harland-Lang, A.~D. Martin, P.~Motylinski and R.~S. Thorne, \emph{{Parton
  distributions in the LHC era: MMHT 2014 PDFs}},
  \href{https://doi.org/10.1140/epjc/s10052-015-3397-6}{\emph{Eur. Phys. J.}
  {\bfseries C75} (2015) 204}
  [\href{https://arxiv.org/abs/1412.3989}{{\ttfamily 1412.3989}}].

\bibitem{Ball:2014uwa}
{\scshape NNPDF} collaboration, R.~D. Ball et~al., \emph{{Parton distributions
  for the LHC Run II}},
  \href{https://doi.org/10.1007/JHEP04(2015)040}{\emph{JHEP} {\bfseries 04}
  (2015) 040} [\href{https://arxiv.org/abs/1410.8849}{{\ttfamily 1410.8849}}].

\bibitem{Gao:2013bia}
J.~Gao and P.~Nadolsky, \emph{{A meta-analysis of parton distribution
  functions}}, \href{https://doi.org/10.1007/JHEP07(2014)035}{\emph{JHEP}
  {\bfseries 07} (2014) 035} [\href{https://arxiv.org/abs/1401.0013}{{\ttfamily
  1401.0013}}].

\bibitem{Carrazza:2015aoa}
S.~Carrazza, S.~Forte, Z.~Kassabov, J.~I. Latorre and J.~Rojo, \emph{{An
  Unbiased Hessian Representation for Monte Carlo PDFs}},
  \href{https://doi.org/10.1140/epjc/s10052-015-3590-7}{\emph{Eur. Phys. J.}
  {\bfseries C75} (2015) 369}
  [\href{https://arxiv.org/abs/1505.06736}{{\ttfamily 1505.06736}}].

\bibitem{Watt:2012tq}
G.~Watt and R.~S. Thorne, \emph{{Study of Monte Carlo approach to experimental
  uncertainty propagation with MSTW 2008 PDFs}},
  \href{https://doi.org/10.1007/JHEP08(2012)052}{\emph{JHEP} {\bfseries 08}
  (2012) 052} [\href{https://arxiv.org/abs/1205.4024}{{\ttfamily 1205.4024}}].

\bibitem{deFlorian:2015ujt}
D.~de~Florian, G.~F.~R. Sborlini and G.~Rodrigo, \emph{{QED corrections to the
  Altarelli-Parisi splitting functions}},
  \href{https://doi.org/10.1140/epjc/s10052-016-4131-8}{\emph{Eur. Phys. J.}
  {\bfseries C76} (2016) 282}
  [\href{https://arxiv.org/abs/1512.00612}{{\ttfamily 1512.00612}}].

\bibitem{Buckley:2014ana}
A.~Buckley, J.~Ferrando, S.~Lloyd, K.~{Nordstr\"om}, B.~Page, M.~{R\"ufenacht}
  et~al., \emph{{LHAPDF6: parton density access in the LHC precision era}},
  \href{https://doi.org/10.1140/epjc/s10052-015-3318-8}{\emph{Eur. Phys. J.}
  {\bfseries C75} (2015) 132}
  [\href{https://arxiv.org/abs/1412.7420}{{\ttfamily 1412.7420}}].

\bibitem{Aaboud:2016yus}
{\scshape ATLAS} collaboration, M.~Aaboud et~al., \emph{{Measurement of the
  $W^{\pm}Z$ boson pair-production cross section in $pp$ collisions at
  $\sqrt{s}=13$ TeV with the ATLAS Detector}},
  \href{https://doi.org/10.1016/j.physletb.2016.08.052}{\emph{Phys. Lett.}
  {\bfseries B762} (2016) 1}
  [\href{https://arxiv.org/abs/1606.04017}{{\ttfamily 1606.04017}}].

\bibitem{Agashe:2014kda}
{\scshape Particle Data Group} collaboration, K.~Olive et~al., \emph{{Review of
  Particle Physics}},
  \href{https://doi.org/10.1088/1674-1137/38/9/090001}{\emph{Chin.Phys.}
  {\bfseries C38} (2014) 090001}.

\bibitem{Khachatryan:2016tgp}
{\scshape CMS} collaboration, V.~Khachatryan et~al., \emph{{Measurement of the
  WZ production cross section in pp collisions at $\sqrt(s) =$ 13 TeV}},
  \href{https://doi.org/10.1016/j.physletb.2017.01.011}{\emph{Phys. Lett.}
  {\bfseries B766} (2017) 268}
  [\href{https://arxiv.org/abs/1607.06943}{{\ttfamily 1607.06943}}].

\end{thebibliography}

\providecommand{\href}[2]{#2}\begingroup\raggedright\endgroup

\end{document}